\documentclass[a4paperal]{i3m}

\usepackage{cite}
\usepackage{amsmath,amssymb,amsfonts}
\usepackage{algorithmic}
\usepackage{graphicx}
\usepackage{textcomp}
\usepackage{xcolor}
\usepackage{bm}
\usepackage{algorithm}
\usepackage{algorithmic}
\usepackage{makecell}
\usepackage{graphicx}
\usepackage{textcomp}
\usepackage{array} 
\usepackage{longtable}
\usepackage{booktabs}
\usepackage{float}
\usepackage{diagbox}
\usepackage{balance}

\conference{emss}

\title{Nonstationary Continuum-Armed Bandit Strategies for Automated Trading in a Simulated Financial Market}

\author[1,2]{Bingde Liu}
\author[1,\authfn{1}]{John Cartlidge}

\affil[1]{Department of Computer Science, University of Bristol, Bristol, BS8 1UB, UK}
\affil[2]{Department of Industrial Engineering and Economics, Tokyo Institute of Technology, Tokyo, 152-8550, Japan}

\authnote{\authfn{1}Corresponding author. Email address: john.cartlidge@bristol.ac.uk}

\runningauthor{B. Liu and J. Cartlidge}

\confyear{2023}

\begin{document}

\begin{frontmatter}
\maketitle
\begin{abstract}
We approach the problem of designing an automated trading strategy that can consistently profit by adapting to changing market conditions. This challenge can be framed as a Nonstationary Continuum-Armed Bandit (NCAB) problem. To solve the NCAB problem, we propose PRBO, a novel trading algorithm that uses Bayesian optimization and a ``bandit-over-bandit'' framework to dynamically adjust strategy parameters in response to market conditions. We use Bristol Stock Exchange (BSE) to simulate financial markets containing heterogeneous populations of automated trading agents and compare PRBO with PRSH, a reference trading strategy that adapts strategy parameters through stochastic hill-climbing. Results show that PRBO generates significantly more profit than PRSH, despite having fewer hyperparameters to tune. The code for PRBO and performing experiments is available online open-source (\url{https://github.com/HarmoniaLeo/PRZI-Bayesian-Optimisation}). 
\end{abstract}

\end{frontmatter}

\section{Introduction}

Automation now pervades most aspects of trading in financial markets \citep{algo2020sec}. 
In equity markets, there is a proliferation of electronic, order-driven trading venues that operate at (or close to) nanosecond timescales; while the majority of orders sent to exchanges are generated by automated execution algorithms that trade autonomously on behalf of investors.  This interconnected network of electronic trading venues populated by automated trading algorithms has resulted in contemporary financial markets that move at lightning-fast speeds and present new forms of systemic risks such as ultra-fast price swings and flash crashes \citep{cartlidge18complex}.

Simulation can be used to model and better understand the complex dynamics of financial markets and there is a long history of {\em agent-based computational economics}, where ``zero-intelligence'' (ZI) agents -- simple rule-based strategies -- are used to model interacting populations of financial traders competing for profit \citep{ladley2012zi}. Recently,  \citet{cliff2021przi} introduced Parameterised-Response Zero Intelligence (PRZI; pronounced ``prezzy''), a ZI agent with a single strategy parameter $s \in [$-$1,1]$ that controls the behaviour of PRZI and enables it to act like several other reference ZI strategies, or some hybrid mix. By altering $s$, it is possible to adapt PRZI to changing market conditions, which emulates the continuous competition for profits that we observe in real markets. To this end, \citet{cliff22prsh} introduced PRZI-Stochastic-Hillclimber (PRSH; pronounced ``purr-sh''), an adaptive trading agent that uses stochastic hillclimbing to autonomously adjust parameter $s$ during a simulated trading session in an attempt to maximise profit generation. It has been shown that PRSH is more profitable than PRZI with some fixed strategy value $s$, and simulated markets containing populations of PRSH agents produce competitive co-adaptive dynamics reminiscent of the real world  \citep{cliff22prsh}.

In this paper, we attempt to introduce an improved algorithm for adapting PRZI strategy $s$. 
Since $s$ is a continuous value (i.e., we have a continuum) and the distribution of its performance metric changes over time (i.e., payoff is nonstationary), we frame the online tuning process as a \emph{Nonstationary Continuum-Armed Bandit} (NCAB) problem. To solve this problem, we propose a new adaptive trading algorithm, which we name PRZI-Bayesian Optimisation (PRBO; pronounced ``purr-boh''). PRBO decomposes the NCAB problem into two, and solves the Continuum Armed-Bandit (CAB) sub-problem by Bayesian optimization, and the Nonstationary Multi-Armed Bandit (NMAB) sub-problem by utilising an adapted version of the ``bandit-over-bandit'' framework, first introduced by \citet{cheung2022hedging}. 

To evaluate PRBO, we use the Bristol Stock Exchange \citep[BSE;][]{cliff2018bse}, a minimal simulation of a centralized financial market based on a continuous double auction running via a limit order book (LOB). We populate markets with a variety of heterogeneous trading agents, and directly compare the performance of PRBO against PRSH in markets with, and without, trends. Results show that PRBO generates significantly more profit than PRSH under all market conditions, whilst also benefiting from having fewer tunable hyperparameters.

\vspace{2mm}
\noindent
{\bf Summary of contributions:}
\begin{enumerate}
    \item We propose PRBO, a new adaptive trading algorithm that has fewer hyperparameters than PRSH. 
    \item We perform empirical evaluations of PRBO and PRSH in simulated financial markets with varying dynamics. 
    \item We demonstrate that the PRBO generates significantly more profit than PRSH. 
\end{enumerate}

\section{Related Works}

In this section, we present existing research related to this work, including existing research on the MAB problem and financial market simulation.

\subsection{MAB Problem: Contextual Background}

In the Multi-Armed Bandit (MAB) problem~\citep{slivkins2019introduction}, 
a gambler makes a series of attempts to pull different arms of a
multi-armed bandit. The payoff for pulling each arm has a
different unknown probability distribution. Given that only a
finite number of attempts can be made, the gambler's objective
is to find a sequence of arm pulls that maximizes reward.
A lightweight online learning algorithm can be used to adjust 
the arm selection strategy, using the payoff received 
from each arm pull as feedback.

Under the basic MAB problem setting, there are a finite number $N$ arms to pull and the payoff distribution for pulling each arm remains constant. Many studies have considered the basic MAB problem \citep[e.g.,][] {slivkins2019introduction,berry1985bandit,russo2018tutorial}. Classical algorithms for solving the basic MAB problem include \emph{Uniform Exploration} and its improvements \emph{Epsilon-Greedy} and \emph{Softmax Epsilon-Greedy}. Later developments 
include the \emph{Upper Confidence Bound (UCB)} algorithm, which makes use of 
payoff confidence intervals, and \emph{Thompson
sampling}~\citep{russo2018tutorial}, which uses a Bayesian generative model. 

There exist more complex variants of the MAB problem. For example, the \emph{Continuum-Armed Bandit} (CAB) problem considers cases where there are an infinite number of arms \citep{agrawal1995continuum}. The CAB problem is often approached by using a disretization algorithm to divide the domain into subintervals such that CAB is effectively converted to MAB with finite $N$ \citep[e.g.,][]{auer2002nonstochastic,kleinberg2004nearly,auer2007improved}, and can include a {\em zooming algorithm} to focus exploration on areas near apparent maxima \citep{kleinberg2008multi}. A related variant of MAB is the {\em Finite Continuum-Armed Bandit} (F-CAB) problem, where an agent is presented with $N$ arms in a continuous space \citep{gaucher20finitecontinuum}.

If the probability distribution of payoff by pulling each arm changes over time, it is considered a \emph{Nonstationary Multi-Armed Bandit} (NMAB) problem. NMAB problems are usually approached by passive adaptive strategies \citep[e.g.,][]{kocsis2006discounted,gonccalves2015upper,besbes2014stochastic}, active adaptive strategies~\citep[e.g.,][]{hartland2007change,mellor2013thompson}, or a mixture of both \citep{allesiardo2015exp3}. However, a recent study by \citet{cheung2022hedging} introduced a novel ``bandit-over-bandit'' framework that adapts to latent changes in payoff distributions and can discover near-optimal solutions to NMAB problems in a surprisingly parameter-free manner. 

\subsection{Financial Market Simulation}

Multi-agent simulations are commonly used to simulate financial markets \citep[e.g.,][]{lux1999scaling,lebaron2001builder,samanidou2007agent} and have been used to investigate various phenomena, such as market microstructure~\citep{muranaga1999market}, market regulation~\citep{mizuta2016brief}, market fragmentation~\citep{duffin18}, and market dynamics~\citep{shi23}, etc. 

In this paper, we perform market simulations using the Bristol Stock Exchange (BSE)~\citep{cliff2018bse}. BSE is a minimal simulation of a centralized financial market based on a continuous double auction running via a limit order book (LOB). It can be populated by automatic trader agents entering the market at a different time with their limit prices, placing quotes on the LOB, and making orders executed as much as possible at a better price to make a profit. BSE includes a selection of reference trading algorithms from the literature, and is available online open-source (\url{https://github.com/davecliff/BristolStockExchange}).

BSE contains a selection of reference trading algorithms from the literature, including: 
Giveaway \citep[GVWY;][]{cliff2018bse}, Zero-Intelligence Constrained \citep[ZIC;][]{gode1993allocative}, Shaver \citep[SHVR;][]{cliff2018bse}, Sniper \citep[SNPR;][]{rust1993behavior}, Zero-Intelligence Plus \citep[ZIP;][]{cli1997minimal}, Parameterised-Response Zero Intelligence \citep[PRZI;][]{cliff2021przi}, and PRZI-Stochastic-Hillclimber \citep[PRSH;][]{cliff22prsh}. 

In this work, we focus attention on PRZI, which has a single strategy parameter $s \in [$-$1,1]$ that controls whether PRZI behaves as a pure SHVR, ZIC, or GVWY strategy, or some hybrid mixture. We aim to introduce a novel algorithm for automatically adapting $s$ that can outperform PRSH (see Section~\ref{sec:bridge} for a detailed introduction).

\section{Technical Background}

Here, we present necessary technical foundations. 

\subsection{MAB Problem: Technical Formulation}

Under the basic MAB problem setting, assume that $\mathrm{A}$ is the set of arms, $a\in \mathrm{A}$ is the arm to pull, and pulling arm $a$ gives payoff $r$ which conforms to a probability distribution $D(a)$ with expectation $\mu(a)$, i.e.,: 
\begin{equation}
 r\sim D(a),\quad \mu(a) = \mathbb{E}(D(a)). 
\end{equation}
\noindent We evaluate bandit algorithms by \emph{regret}, which is the difference between the current theoretical optimal reward and the current reward. Assume there is an optimal arm $a^*$ to pull which gives best expected reward $\mu^*=\max\limits_{\alpha \in A}\mu(a)=\mu(a^*)$, then the regret at time $t$ is defined as $R(t):=\mu^*t-\sum\limits_{s=1}^t\mu(a_s)$. The objective of the MAB algorithms is to minimize the regret over the process. To solve the problem, MAB algorithms aim to balance \emph{exploitation} and \emph{exploration}~\citep{slivkins2019introduction}.

During the whole process, assume that $a_t\in \mathrm{A},t \in \{1,...,T\}$, where $t \in \{1,...,T\}$ are timestamps to pull the arms. In the \emph{Continuum-Armed Bandit} problem \citep{agrawal1995continuum}, all arms are considered forming an infinite set $\mathrm{A}$ satisfying $a\in[0,1],\forall a\in\mathrm{A}$. Then the average reward satisfies the Lipschitz continuum:
\begin{equation}
    |\mu(x)-\mu(y)|\leq L|x-y|,\quad\forall x,y\in \mathrm{A},
\end{equation}
\noindent On the other hand, in the \emph{Nonstationary Continuum-Armed Bandit}  problem, $D(a)$ migrate slowly with time. Therefore, the payoff of pulling each arm $R(a)$ conforms to a stochastic process $D_t(a)$. 

\subsection{Trading Strategy Selection as a MAB Problem}

PRZI contains a single parameter, $s$, which determines strategy behaviour.  We can evaluate the performance of a particular \emph{s} value by \emph{profit-per-second} (\emph{pps}), which is determined by the profit made by a single transaction divided by the time that a specific value of \emph{s} exists. Under a certain value of \emph{s}, the greater the value of \emph{pps}, the better the \emph{s} value. Since traders who receive limit orders at different prices enter the market at random, \emph{pps} is affected by the uncertainties in the market, i.e., it is a stochastic process. At the same time, the market is often dynamic, with a changing demand and supply range, so the probability distribution that \emph{pps} conforms to is also changing with time. Therefore, \emph{pps} conforms to a stochastic process consisting of a cluster of random variables related to \emph{s} and \emph{t}, denoted as $D_t(s)$. 

In the whole transaction process, the value of $s$ can be changed at any time. Since each trader can execute transactions finite times, the chances to change the value of $s$ are limited. Therefore, we can consider the online parameter tuning problem of the PRZI algorithm by regarding $s$ as the arms of a bandit in the MAB problem, with $pps$ as the payoff. Furthermore, since $s$ is a continuous value, and the distribution of $pps$ changes over time, it is both a Continuum-Armed Bandit problem and a Nonstationary Bandit problem. Together, we call it a \emph{Nonstationary Continuum-Armed Bandit (NCAB)} problem. 

We group the market ticks $t\in\{1,2,...,T\}$ in BSE to a set of stages $\rho\in\{0,1,... ,P\}$. The tuning of $s$ will be performed at each stage. 

\subsection{PRSH Trading Agent}
\label{sec:bridge}

PRSH is an adaptive version of PRZI, which uses a k-point stochastic hill climber to adapt its value of $s$ over time.  At the stage $\rho$, the PRSH algorithm creates a finite set $S_{\rho}\in [-1,1]$ of $s$. Each $s$ is tried $N$ times, from which the $s$ with the highest average $pps$ is selected and denoted as $s_{\rho}$. Next, the PRSH algorithm will generate $S_{\rho+1}=M(s_{\rho})$ by a function $M$. Usually, $M$ generates $S_{\rho+1}$ by sampling from a normal distribution $N(s_{\rho},\sigma^2)$ with $s_{\rho}$ as expectation and a specific variance $\sigma^2$. 

Assuming $\mu_{\rho}(x):=\mathbb{E}(D_{\rho}(x))$ is continuous on $s$ and varies slowly with stage, satisfying Lipschitz continuum: \begin{equation}|\mu_{\rho}(x)-\mu_{\rho+1}(y)|\leq L|x-y|,\quad\forall x,y\in [-1,1],\end{equation}When $D_{\rho}(s)$ is constant with stage, according to the idea of zooming~\citep{kleinberg2008multi}, $M$ should sample $s$ from a distribution whose variance decreases with stage, to allow $\underset{\rho\rightarrow \infty}{\lim} P(s^*\in S_{\rho})=1$. However, when $D_{\rho}(s)$ is changing with stage, which means that $s^*$ is also changing with stage, the zooming idea will fail. $P(s_{\rho}^*\in S_{\rho})$ may be a quantity that does not converge as $\rho$ increases. In PRSH, $M$ is a pre-determined function that is constant with time, which makes the PRSH algorithm unable to handle NCAB problems and increases the cost of hyperparameter selection. 

There are also other hyperparameters in PRSH: \emph{k} represents the number of $s$ in $S_{\rho}$ and \emph{N} is the number of attempts per \emph{s}. Since the trader-agents will receive orders randomly at intervals, the total number of quotes placed by a single trader agent is uncertain throughout the trading process. Therefore, $N$ is replaced by a time window $W$. Each stage $\rho\in\{1,2,... ,P\}$ will contain $k\times W$ ticks $t$. The total number of phases will be $P=\lceil \frac{T}{k\times W}\rceil$. \emph{Strategy wait time}, noted as $v$, will determine $W$ together with $T$ and $k$, i.e. $W=\lfloor\frac{v}{k}\rfloor$. In summary, in the PRSH algorithm, we have three hyperparameters $k$, $v$, and $M$ that need to be determined. 

\section{PRZI-Bayesian-Optimization (PRBO)}

To address the shortcomings of the existing PRSH algorithm, we propose the PRZI-Bayesian-Optimization (PRBO) algorithm, which uses a Bayesian optimization approach to solve the Continuum Bandits problem and the ``bandit-over-bandit'' framework to solve the nonstationary Bandits problem. In this section, PRBO is introduced in detail, and full pseudocode is presented in Algorithm~\ref{alg:prb}.

\subsection{Bayesian Optimization}
\label{Bayesian optimization}

The Bayesian optimization algorithm~\citep{snoek2012practical,brochu2010tutorial} is based on the Gaussian process. Assume that \emph{pps} is a black-box function $f(s)$ with \emph{s} as the independent variable, that is a realization of a Gaussian process (GP) with mean function $\mu(s)$ and Gaussian kernel covariance function $k(s,s')$, i.e., $f(s)\sim\mathcal{GP}\left(\mu(s),k(s,s')\right)$. In each stage, at each tick when the agent is chosen for trading, a GP regression model will be built based on the observed data $D_t=\{(s_i,y_i)\}_{i=1}^n$, where $y_i=f(s_i)+\epsilon_i$ and $\epsilon_i$ is Gaussian noise with zero mean and variance $\sigma_n^2$. Assume that $S^{*}=\{s^*_i\}_{i=1}^{n^*}$ refers to the set of possibly optimized unexplored $s$ at which the function values is to be predicted using the GP posterior distribution. In contrast, $S_t=\{s_i\}_{i=1}^n$ refers to the set of explored $s$. Conditioned on the conditions mentioned above, the posterior distribution over the latent function values $f$ can be computed analytically using Bayes' rule as follows:
\begin{equation}
  p(f_*|S_*,S_t,D_t) = \mathcal{N}(f_*|\mu_*,\Sigma_*),  
\end{equation}
where $\mu_*=\mu(S_*) + K(S_*,S_t)[K(S_t,S_t)+\sigma_n^2I]^{-1}(y-\mu(S_t))$ and $\Sigma_*=K(S_*,S_*) - K(S_*,S_t)[K(S_t,S_t)+\sigma_n^2I]^{-1}K(S_t,S_*)$ are the predictive mean and covariance matrix, respectively, and $K(S,S')=[k(s,s')]_{s\in S,s'\in S'}$ is the Gram matrix of pairwise kernel evaluations between inputs.

The expected improvement (EI) acquisition function measures the utility or potential benefit of evaluating the function $f$ at a new point $s_{t+1}$,  defined as follows:
\begin{equation}
\mathrm{EI}(s)=\begin{cases}
0 & \text{if }\sigma^*(s)\leq 0\\
(\mu^*(s)-f(s_{\mathrm{best}}))\Phi(Z) + \sigma^*(s)\phi(Z) & \text{otherwise},
\end{cases}
\end{equation}
\noindent
where $\mu_*(s)$ and $\sigma_*(s)$ are the predictive mean and standard deviation at $s$, respectively, $f(s_{\mathrm{best}})$ is the best observed function value so far, $\Phi$ and $\phi$ are the CDF and PDF of the standard normal distribution, respectively, and $Z=(\mu^*(x)-f(x_{\mathrm{best}}))/\sigma_*(x)$ is the standardized improvement. Intuitively, this acquisition function balances exploration (sampling uncertain regions) and exploitation (sampling promising regions) by favoring regions with high predicted mean and/or high predictive uncertainty.

The optimization problem then becomes searching the next $s$ to evaluate that maximizes the acquisition function, i.e., $s_{t+1}=\arg\max_{s\in\mathcal{[-1,1]}}\mathrm{EI}(s)$. The searching process will continuously perform at each stage $\rho\in\{0,1,...,P\}$. 

\begin{algorithm}[tb]
	%\textsl{}\setstretch{1.8}
	\renewcommand{\algorithmicrequire}{\textbf{Input:}}
	\renewcommand{\algorithmicensure}{\textbf{Output:}}
	\caption{PRBO: PRZI-Bayesian-Optimization Strategy} 
	\label{alg:prb}
	\begin{algorithmic}[1]
		\STATE Initialization: 
		\STATE $G_{\rho} := \{g_{\rho,i}\},~ i=1,2,... ,k$
		\STATE $t\leftarrow 1$
		\STATE $\rho\leftarrow 1$
		\STATE $R_i\leftarrow 0, i=1,2,... ,k$
		\STATE $n_i\leftarrow 0, i=1,2,... ,k$
		\FOR{$\rho\in \{1,2,...,P\}$}
		\FOR{$i\in \{1,2,...,k\}$} 
		\FOR{$t\in \{(p-1)\times k\times W,(p-1)\times k\times W+1,...,p\times k\times W\}$}
		\IF{agent chosen to place an order}
		\STATE Using $g_{\rho,i}$ to sample a $s$
		\STATE Using $s$ to bid or ask 
		\STATE $n_i\leftarrow n_i+1$
		\STATE $t_{buf}\leftarrow t$
		\ENDIF
		\IF{the order is executed}
		\STATE Get reward $r$, i.e., profit
		\STATE $R_i\leftarrow R_i+r$
		\FOR{$j\in \{1,2,...,k\}$}
		\STATE Update $g_{\rho,j}$ with the pair $s,r/(t-t_{buf})$
		\ENDFOR
		\ENDIF
		\ENDFOR
		\ENDFOR
		\STATE $\bar\mu_i=R_i/n_i, i=1,2,... ,k$
		\STATE Sample $k-1$ samples of $g$ with $p(g_{\rho,i})=e^{\bar\mu_i}/\sum\limits_{i\in \{1,2,...,k\}}e^{\bar\mu_i}$ without replacement and discard the remaining one $g$
		\STATE Generate a new $g$, forming $G_{\rho+1}$ together with the $(k-1)$ $g$, above
		\STATE $R_i\leftarrow 0, i=1,2,... ,k$
		\STATE $n_i\leftarrow 0, i=1,2,... ,k$
		\ENDFOR 
	\end{algorithmic}
\end{algorithm}

\subsection{Bandit-Over-Bandit Framework}

The ``bandit-over-bandit'' framework was recently introduced by \citet{cheung2022hedging} to adapt to latent changes in the environment. It works by dividing the time horizon into multiple blocks and treating each block as a separate bandit problem, using a bandit algorithm (called the {\em slave algorithm}) to solve it. Another bandit algorithm (called the {\em meta-algorithm}) is applied to tune the slave algorithm at the end of each temporal block. It also uses a ``forgetting principle'' in the learning process, which gives less weight to older data as time goes on and is vital in changing environments. The framework allows the algorithm to enjoy nearly optimal dynamic regret bounds in a parameter-free manner. We leverage the time horizon division and the ``forgetting principle'' proposed in the original work and adapt it to fit the BSE problem formulation. 
Algorithm~\ref{alg:prb} presents our PRBO trading agent implementation of the bandit-over-bandit strategy. Lines 8-24 describe the slave algorithm; lines 7 and 25-30 describe the meta-algorithm.

We use the stages $\rho\in\{0,1,\ldots,P\}$ as the time horizon division. In the original work, \citet{cheung2022hedging} use the sliding window-upper confidence bound algorithm as the slave algorithm. In our work, to solve the Continuum-Armed  Bandit problem, we use Bayesian optimization as the slave algorithm (described in Section~\ref{Bayesian optimization}). 

In order to tune the slave Bayesian optimization algorithm, we propose a novel meta-algorithm (Algorithm~\ref{alg:prb}; lines 7, 25-30). Note that $\bm{s}_t$ and $\bm{y}_t$ respectfully record all explored $s$ and their corresponding payoffs. However, if the environment changes, the posterior distribution obtained from the observations so far may become inaccurate. In this case, according to the ``forgetting principle'' \citep{cheung2022hedging}, rather than continuing to adjust the distribution on the current basis, it would be better to abandon all previous observations and start from scratch. Ideally, we would maintain several different Gaussian processes, each starting to observe $s$ and making an adjustment at a different time, i.e., having different lengths of memory. 

To achieve this, we maintain $k$ Gaussian processes simultaneously and use the Softmax Epsilon-Greedy algorithm \citep{slivkins2019introduction} to selectively drop the observations of certain Gaussian processes at each stage. Let  each Gaussian process be $g_i, i=1,2,... ,k$, and the set of Gaussian processes be $\mathrm{G}=\{g_i\}_{i\in\{1,2,... ,k\}}$. $W$, $\rho$ determined by means in Section~\ref{sec:bridge}. Then the flow of our algorithm is shown in Algorithm~\ref{alg:prb}. 

Using the Softmax Epsilon-Greedy algorithm to randomly drop a Gaussian process at each stage, we can obtain $k$ Gaussian processes with different memory lengths after enough stages have been performed. 

Compared to PRSH, PRBO has only two hyperparameters, $k$ and $v$, making its hyperparameter selection less time costly. In the subsequent experiments, we will compare PRBO with PRSH to determine which has the best profit-maximising performance. 

\section{Experiment Design}

In this section, we present our experimental design. We will first introduce the setup of the market simulation, then introduce the method of hyperparameter selection for PRSH and PRBO, and finally introduce the experiment used to compare the performance of PRSH and PRBO.

\subsection{Market Simulation Method}

\begin{figure}[tb]
\centering
\includegraphics[width=0.8\linewidth]{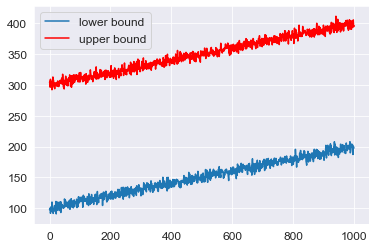}
\caption{Supply and demand range of a trending market, $e_{trend}$.}
\label{fig:trend-market}
\end{figure}

\begin{figure}[tb]
\centering
\includegraphics[width=0.8\linewidth]{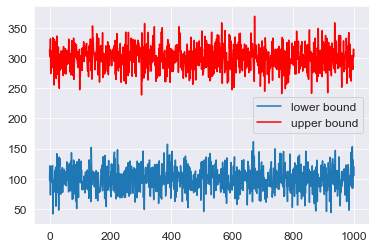}
\caption{Supply and demand range of a flat market, $e_{flat}$.}
\label{fig:trendless-market}
\end{figure}

We use BSE to generate experimental data with 1000 seconds simulation. New orders arrive at intervals modeled with a Poisson distribution, like a real market. We generate symmetrical supply-demand curves, but supply and demand ranges are changing over time. According to different market dynamics, the ways of change are also different. For trending markets, our supply and demand range is $[0.1 \times t+N(0,5)+100,0.1 \times t+N(0,5)+300]$ as shown in Fig.~\ref{fig:trend-market}. For markets without trend, our supply and demand range is $[N(0,20)+100,N(0,20)+300]$ as shown in Fig.~\ref{fig:trendless-market}. In both figures, red (blue) lines represent the upper (lower) limits of the supply and demand ranges. The $N(\mu,\sigma)$ indicates a white Gaussian noise with mean $\mu$ and standard deviation $\sigma$. Therefore, we simulate: (i) a {\em trending market} in which the supply and demand range increases linearly with time and has relatively low volatility; and (ii) a {\em flat market} in which the supply and demand range does not change with time but has greater volatility. We represent market dynamics as $e\in\{e_{trend},e_{flat}\}$. 

To emulate more realistic dynamics, we populate markets with a heterogeneous variety of different trading agent strategies. When performing hyperparameters selection, both buyers and sellers are 20 GVWY traders, 20 ZIC traders, 20 ZIP traders, 20 SNPR traders, 20 SHVR traders, and 20 traders with algorithms either PRSH or PRBO. When comparing the performance of the two algorithms, we include both traders using PRSH and PRBO, which means both buyers and sellers are 20 GVWY traders, 20 ZIC traders, 20 ZIP traders, and 20 SNPR traders, 20 SHVR traders, 20 PRSH traders, and 20 PRBO traders. 

\subsection{Hyperparameters Exploration Method}
%\subsubsection{PRSH}
For PRSH, we consider $k \in \{2,4,6,\ldots,16\}$, $v \in \{32,64,128,256\}$, and three mutation functions, $m\in\{m_1,m_2,m_3\}$: 
\begin{description}
    \item[m$_1$: ] $s_i = M(s) = s_0+N(0,0.05), i=1,2,...,k$
    \item[m$_2$: ] $s_i = M(s) = s_0+N(0,0.15), i=1,2,...,k$
    \item[m$_3$: ] $s_i = M(s,i) = \begin{cases}s_0+\mathbb{U}(0,0.1), i=1,3,...,k/2-1\\s_0-\mathbb{U}(0,0.1), i=2,4,...,k/2\end{cases}$
\end{description}
We repeat 100 experiments in each market dynamic $e$ with each combination of parameters, and record the total profit per PRSH trader per experiment as a sample $x^i_{e,k,v,m},i=1,2,...,100$. Note the i.i.d. samples as $\bm{x_{e,k,v,m}}=(x^1_{e,k,v,m},x^2_{e,k,v,m},...,x^{100}_{e,k,v,m})$, which are sampled from a distribution $X(e,k,v,m)$. We will estimate $\mathbb{E}[X(e,k,v,m)]$ by $\hat{\mathbb{E}}[X(e,k,v,m)]=\bar{\bm{x_{e,k,v,m}}}$, and observe the distribution of $\bm{x_{e,k,v,m}}$ for different $k,v,m$. Then we attempt to find the best combination  $k^*,v^*,m^*=\arg\max\limits_{k,v,m*}\hat{\mathbb{E}}[X(e,k,v,m)]$, which is the possible optimal parameter obtained from the sample by estimating $\mathbb{E}[X(e,k^*,v^*,m^*)]$. 

For PRBO, We explore $k \in \{2,3,4\}$ and $v \in \{32,64,128,256\}$. We repeated 100 experiments in each market dynamic $e$ with each combination of parameters, and record the total profit per PRBO trader made in each experiment as a sample $y^i_{e,k,v},i=1,2,...,100$. Note the i.i.d. samples as $\bm{y_{e,k,v}}=(y^1_{e,k,v},y^2_{e,k,v},...,y^{100}_{e,k,v})$, which are sampled from a distribution $Y(e,k,v)$. We will estimate $\mathbb{E}[Y(e,k,v)]$ by $\hat{\mathbb{E}}[Y(e,k,v)]=\bar{\bm{y_{e,k,v}}}$, and observe the distribution of $\bm{y_{e,k,v}}$ for different $k,v$. Then we are going to find the best combination $k^*,v^*=\arg\max\limits_{k,v}\hat{\mathbb{E}}[Y(e,k,v)]$, this is the possible optimal parameter obtained from the sample by estimating $\mathbb{E}[Y(e,k^*,v^*)]$. 

To prove the optimality of the parameters we obtained, we need to perform hypothesis testing. We first test the normality of $X(e,k,v,m)$ and $Y(e,k,v)$ using the Kolmogorov–Smirnov test (K-S test). If $X(e,k,v,m)$ and $Y(e,k,v)$ conform to the normal distribution we can then perform Z-test to test whether $\mathbb{E}[X(e,k^*,v^*,m^*)]>\mathbb{E}[X(e,k,v,m)],\forall k,v,m$ and $\mathbb{E}[Y(e,k^*,v^*)]>\mathbb{E}[Y(e,k,v)],\forall k,v$. We will perform Z-test on every $\bm{x_{e,k,v,m}}$ and $\bm{y_{e,k,v}}$ that we have obtained one-by-one with $\bm{x_{e,k^*,v^*,m^*}}$ and $\bm{y_{e,k^*,v^*}}$ respectively. We record all combinations of parameters that do not significantly make less profit than the best combinations. Eventually, all recorded parameter combinations, together with the best combinations, will form $\{K_X^*(e),V_X^*(e),M_X^*(e)\}$ 
(for PRSH) and $\{K_Y^*(e),V_Y^*(e)\}$ (for PRBO). 

\subsection{PRBO vs PRSH: Comparison Experiment Design}

To compare the performance of PRSH traders and PRBO traders, we put both traders into the market. Since $\{K_X(e)^*,V_X(e)^*,M_X(e)^*\}$ and $\{K_Y(e)^*,V_Y(e)^*\}$ contain the possibly optimal hyperparameter combinations for the PRSH and PRBO respectively, each PRSH trader will randomly choose $\{k,v,m\}$ from $\{K_X(e)^*,V_X(e)^*,M_X(e)^*\}$ and each PRBO trader will randomly choose $\{k,v\}$ from $\{K_Y(e)^*,V_Y(e)^*\}$. 

We repeated 100 experiments in each market dynamic $e$. In each experiment, we record the difference between the total profit per PRBO trader and the total profit per PRSH trader as a sample $d^i_e,i=1,2,...,100$. Note the i.i.d. samples as $\bm{d_e} = (d^1_e,d^2_e,...,d^{100}_e)$, which are sampled from a distribution $D(e)$. We will estimate $\mathbb{E}[D(e)]$ by $\hat{\mathbb{E}}[D(e)]=\bar{\bm{d_e}}$. What we will be interested in is whether $\mathbb{E}[D(e)]>0$. If we can show by hypothesis testing that $\mathbb{E}[D(e)]>0$, then we have good reason to believe that PRBO outperforms PRSH. In this case, we test the normality of $D(e)$ using the K-S test and then perform Z-test to test whether $\mathbb{E}[D(e)]>0$. 

\begin{table*}[!ht]
    \caption{Mean profit of PRSH traders in trending markets. The highest profit is shown in parentheses. Profits with no underlining are significantly lower than the maximum (Z-test; $p<0.05$). Profits underlined are {\em not} significantly lower than the maximum profit (Z-test; $p$ values shown in subscript).}\label{tab:profit-trending}
    \centering
    \begin{tabular}{c|cccc|cccc|cccc}
    \toprule[1.5pt]%第一条粗线
        M & \multicolumn{4}{c|}{m1}  & \multicolumn{4}{c|}{m2} & \multicolumn{4}{c}{m3} \\ \Xhline{1.2pt}%第一条粗线
        \diagbox{K}{V} & 32 & 64 & 128 & 256 & 32 & 64 & 128 & 256 & 32 & 64 & 128 & 256 \\ \Xhline{1.2pt}
        2 & 1.15 & 1.17 & 1.19 & $\underline{1.24}_{0.10}$ & 1.16 & 1.2 & 1.2 & $\underline{1.24}_{0.14}$ & 1.2 & 1.18 & 1.2 & $\underline{1.23}_{0.11}$ \\ \hline 4 & 1.22 & $\underline{1.27}_{0.39}$ & $\underline{1.23}_{0.08}$ & 1.17 & 1.2 & 1.18 & 1.22 & $\underline{1.22}_{0.06}$ & 1.21 & 1.19 & 1.21 & $\underline{1.23}_{0.09}$ \\ \hline 6 & $\underline{1.26}_{0.28}$ & 1.21 & 1.17 & 1.19 & 1.16 & 1.2 & 1.18 & 1.22 & 1.19 & 1.21 & \textbf{(1.28)} & 1.19 \\ \hline 8 & 1.22 & 1.19 & 1.17 & 1.2 & 1.2 & 1.22 & 1.2 & 1.17 & 1.17 & 1.18 & 1.21 & 1.19 \\ \hline 10 & $\underline{1.23}_{0.12}$ & 1.15 & 1.19 & 1.2 & 1.18 & 1.2 & 1.19 & 1.17 & 1.18 & 1.2 & 1.21 & $\underline{1.23}_{0.08}$ \\ \hline 12 & 1.21 & 1.22 & 1.21 & $\underline{1.27}_{0.38}$ & 1.2 & 1.18 & $\underline{1.26}_{0.36}$ & $\underline{1.25}_{0.20}$ & 1.19 & 1.18 & 1.19 & $\underline{1.24}_{0.12}$ \\ \hline 14 & 1.2 & 1.2 & 1.17 & $\underline{1.25}_{0.19}$ & 1.16 & $\underline{1.24}_{0.15}$ & 1.19 & 1.2 & 1.18 & $\underline{1.22}_{0.05}$ & $\underline{1.23}_{0.07}$ & 1.2 \\ \hline 16 & 1.2 & $\underline{1.25}_{0.19}$ & 1.2 & $\underline{1.27}_{0.48}$ & 1.17 & 1.18 & 1.18 & 1.2 & $\underline{1.24}_{0.14}$ & 1.2 & 1.18 & $\underline{1.23}_{0.10}$ \\ \bottomrule[1.5pt]
    \end{tabular}
\end{table*}

\begin{table*}[!ht]
    \caption{Mean profit of PRSH traders in flat markets. The highest profit is shown in parentheses. Profits with no underlining are significantly lower than the maximum (Z-test; $p<0.05$). Profits underlined are {\em not} significantly lower than the maximum profit (Z-test; $p$ values shown in subscript).}\label{tab:profit-trendless}
    \centering
    \begin{tabular}{c|cccc|cccc|cccc}
    \toprule[1.5pt]%第一条粗线
        M & \multicolumn{4}{c|}{m1}  & \multicolumn{4}{c|}{m2} & \multicolumn{4}{c}{m3} \\ \Xhline{1.2pt}%第一条粗线
        \diagbox{K}{V} & 32 & 64 & 128 & 256 & 32 & 64 & 128 & 256 & 32 & 64 & 128 & 256 \\ \Xhline{1.2pt}
        2 & 1.18 & $\underline{1.23}_{0.09}$ & 1.19 & $\underline{1.23}_{0.08}$ & 1.16 & $\underline{1.23}_{0.06}$ & 1.19 & $\underline{1.27}_{0.46}$ & $\underline{1.23}_{0.11}$ & 1.18 & 1.21 & $\underline{1.23}_{0.06}$ \\ \hline 4 & $\underline{1.24}_{0.11}$ & 1.21 & $\underline{1.25}_{0.19}$ & $\underline{1.27}_{0.38}$ & 1.17 & 1.19 & 1.22 & $\underline{1.27}_{0.37}$ & 1.22 & 1.21 & $\underline{1.26}_{0.31}$ & 1.21 \\ \hline 6 & $\underline{1.23}_{0.10}$ & 1.21 & 1.22 & 1.22 & 1.2 & 1.22 & $\underline{1.26}_{0.37}$ & $\underline{1.24}_{0.09}$ & $\underline{1.23}_{0.10}$ & $\underline{1.25}_{0.16}$ & \textbf{(1.27)} & $\underline{1.24}_{0.10}$ \\ \hline 8 & $\underline{1.25}_{0.21}$ & $\underline{1.26}_{0.30}$ & 1.2 & 1.21 & 1.16 & 1.21 & 1.21 & 1.19 & 1.19 & $\underline{1.25}_{0.23}$ & 1.22 & 1.19 \\ \hline 10 & $\underline{1.24}_{0.12}$ & $\underline{1.22}_{0.05}$ & 1.19 & $\underline{1.24}_{0.13}$ & $\underline{1.25}_{0.18}$ & 1.21 & $\underline{1.22}_{0.06}$ & 1.22 & 1.17 & 1.22 & $\underline{1.24}_{0.17}$ & $\underline{1.24}_{0.15}$ \\ \hline 12 & 1.19 & $\underline{1.25}_{0.19}$ & 1.21 & $\underline{1.23}_{0.10}$ & 1.14 & 1.2 & 1.17 & 1.22 & 1.2 & 1.21 & 1.21 & 1.22 \\ \hline 14 & 1.21 & 1.21 & 1.2 & 1.19 & $\underline{1.23}_{0.10}$ & 1.16 & $\underline{1.27}_{0.48}$ & $\underline{1.24}_{0.13}$ & 1.19 & 1.21 & $\underline{1.24}_{0.17}$ & $\underline{1.26}_{0.35}$ \\ \hline 16 & 1.22 & 1.21 & 1.18 & $\underline{1.23}_{0.06}$ & 1.22 & 1.21 & 1.18 & $\underline{1.24}_{0.15}$ & 1.2 & 1.22 & 1.2 & 1.21 \\ \bottomrule[1.5pt]
    \end{tabular}
\end{table*}

All hypothesis tests in our work will take the significance level $\alpha=0.05$. A detailed description of all hypothesis tests and results are presented in the Appendix. 

\section{Experiment Results}

In this section, we analyze the experimental results. We will show the results of hyperparameter selection, to determine the set of hyperparameters to be used for the comparison experiments. Then we will show the results of the comparison experiments, which demonstrate the superiority of PRBO over PRSH in terms of profit generation.

\subsection{Results of Hyperparameter Selection}

\subsubsection{Optimal PRSH Hyperparameters}

We first perform a K-S test, which demonstrates that $X(e,k,v,m),\forall e,k,v,m$ conforms to the normal distribution (see Appendix for full details). This enables us to use Z-test for statistical comparison of profits.

Table~\ref{tab:profit-trending} and Table~\ref{tab:profit-trendless} shows the mean profit (i.e. $\hat{\mathbb{E}}[X(e,k,v,m)]$) made by PRSH traders under trending market and flat market, respectively. All the $\hat{\mathbb{E}}[X(e,k,v,m)]$ are divided by 1,000 for clarity. The parameters combination with the highest mean profit in the trending market is $(k^*,v^*,m^*)=\arg\max\limits_{k,v}\hat{\mathbb{E}}[X(e_{trend},k,v,m)]=(6,128,m_3)$ with $\hat{\mathbb{E}}[X(e_{trend},k^*,v^*,m^*)]=1277.68$, while the combination with the highest mean profit under flat market is $(k^*,v^*,m^*)=\arg\max\limits_{k,v}\hat{\mathbb{E}}[X(e_{flat},k,v,m)]=(6,128,m_3)$ with $\hat{\mathbb{E}}[X(e_{flat},k^*,v^*,m^*)]=1274.51$. The combinations with the highest mean profit are displayed in bold in the table. 

Under $e_{trend}$, of the full 96 combinations of $k,v,m$, 23 combinations could not reject the null hypothesis in Z-test at the significant level of $\alpha=0.05$ (including $k^*,v^*,m^*$ itself), while 37 combinations under $e_{flat}$. All the combinations are underlined in the table, and the p-values of the Z-test are displayed in the subscript. We will use those combinations of parameters to create $K_X(e_{trend})^*,V_X(e_{trend})^*,M_X(e_{trend})^*$ and $K_X(e_{flat})^*,V_X(e_{flat})^*,M_X(e_{flat})^*$ respectively.

\subsubsection{Optimal PRBO Hyperparameters}

\begin{table}[tb]
    \caption{Mean profit of PRBO traders in trending markets.}\label{tab:trending-profit}
    \centering
    \begin{tabular}{c|cccc}
    \toprule[1.5pt]%第一条粗线
        \diagbox{K}{V} & 32 & 64 & 128 & 256 \\ \Xhline{1.2pt}
        2 & $\underline{2.20}_{0.08}$ & $\underline{2.21}_{0.09}$ & 2.19 & $\underline{2.23}_{0.18}$ \\ \hline 3 & $\underline{2.19}_{0.06}$ & $\underline{2.27}_{0.46}$ & $\underline{2.22}_{0.15}$ & $\underline{2.23}_{0.21}$ \\ \hline 4 & \textbf{(2.28)} & $\underline{2.21}_{0.11}$ & 2.19 & $\underline{2.27}_{0.48}$ \\ \bottomrule[1.5pt]
    \end{tabular}
\end{table}

\begin{table}[tb]
    \caption{Mean profit of PRBO traders in flat markets.}\label{tab:trendless-profit}
    \centering
    \begin{tabular}{c|cccc}
    \toprule[1.5pt]%第一条粗线
        \diagbox{K}{V} & 32 & 64 & 128 & 256 \\ \Xhline{1.2pt}
        2 & \textbf{(2.35)} & $\underline{2.23}_{0.18}$ & $\underline{2.26}_{0.33}$ & $\underline{2.32}_{0.19}$ \\ \hline 3 & $\underline{2.31}_{0.29}$ & $\underline{2.30}_{0.36}$ & $\underline{2.32}_{0.19}$ & $\underline{2.28}_{0.48}$ \\ \hline 4 & $\underline{2.27}_{0.41}$ & $\underline{2.32}_{0.20}$ & $\underline{2.32}_{0.23}$ & $\underline{2.30}_{0.30}$ \\ \bottomrule[1.5pt]
    \end{tabular}
\end{table}

We first perform a K-S test, which demonstrates that $Y(e,k,v),\forall e,k,v$ conforms to the normal distribution (see Appendix for full details). This enables us to use Z-test for statistical comparison of profits.

Table~\ref{tab:trending-profit} and Table~\ref{tab:trendless-profit} show the mean profit (i.e. $\hat{\mathbb{E}}[Y(e,k,v)]$) made by PRBO traders under trending market and flat market respectively. All the $\hat{\mathbb{E}}[Y(e,k,v)]$ are divided by 1,000 for clarity. The parameters combination with the highest mean profit in the trending market is $(k^*,v^*)=\arg\max\limits_{k,v}\hat{\mathbb{E}}[Y(e_{trend},k,v)]=(4,32)$ with $\hat{\mathbb{E}}[Y(e_{trend},k^*,v^*)]=2276.37$, while the combination with the highest mean profit under flat market is $(k^*,v^*)=\arg\max\limits_{k,v}\hat{\mathbb{E}}[Y(e_{flat},k,v)]=(2,32)$ with $\hat{\mathbb{E}}[Y(e_{flat},k^*,v^*)]=2352.83$. The combinations with the highest mean profit are displayed in bold in the table. 

Under $e_{trend}$, of the full 12 combinations of $k,v$, 10 combinations could not reject the null hypothesis in Z-test at the significant level of $\alpha=0.05$ (including $k^*,v^*$ itself), while all 12 combinations under $e_{flat}$. All the combinations are underlined in the table, and the p-values of the Z-test are displayed in the subscript. We will use those combinations of parameters to create $K_Y(e_{trend})^*,V_Y(e_{trend})^*$ and $K_Y(e_{flat})^*,V_Y(e_{flat})^*$ respectively. 

\subsection{Profits Comparison: PRSH vs PRBO}

\begin{figure}[tbp]
\centering
\includegraphics[width=\linewidth]{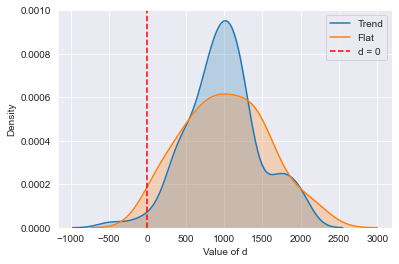}
\label{fig_22}
\caption{Kernel density plot of $\bm{d_{e_{trend}}}$ and $\bm{d_{e_{flat}}}$ showing profit-generating performance advantages of PRBO over PRSH.}\label{fig:kde}
\end{figure}

The kernel density plot of $\bm{d_{e_{trend}}}$ and $\bm{d_{e_{flat}}}$ is shown in Fig.~\ref{fig:kde}. It can be seen from the graph that for both $\hat{\mathbb{E}}[D(e_{trend})]$ and $\hat{\mathbb{E}}[D(e_{flat})]$ the distributions fall largely to the right of the equality line $d=0$. This demonstrates that profits of PRBO are larger than profits of PRSH in both market types.

Table~\ref{statistics} shows the statistic result of $\bm{d_{e_{trend}}}$ and $\bm{d_{e_{flat}}}$. We first perform a K-S test, which demonstrates that both $D(e_{trend})$ and $D(e_{flat})$ pass the normality test at the significance level $\alpha=0.05$ (i.e., K-S test p-values shown in Table~\ref{statistics} are greater than $0.05$). A Z-test is then performed, with p-values of 0.0 showing that the null hypothesis is significantly rejected and $\mathbb E[D(e)]>0,\forall e$. Therefore, we conclude that, on average, the PRBO trading algorithm makes significantly more profit than PRSH trading in both a trending market and a flat market. We present this as strong evidence that PRBO outperforms PRSH. 

\begin{table}[tb]
    %\caption{Delta profit (PRBO - PRSH) in trending and flat markets.}
    \caption{Statistic results of $d_{e_{trend}}$ and $d_{e_{flat}}$ showing PRBO generates significantly higher profits than PRSH in both markets.}
    \label{statistics}
    \centering
    \begin{tabular}{c|cccc}
    \toprule[1.5pt]%第一条粗线
        e & Mean & Std & K-S test p-value & Z-test p-value \\ \Xhline{1.2pt}
        trend & 995.20 & 475.24 & 0.50 & $1.23\times 10^{-98}$ \\
        flat & 1022.44& 550.39 & 0.98 & $4.33\times 10^{-78}$ \\ \bottomrule[1.5pt]
    \end{tabular}
\end{table}

\section{Conclusions}
\label{sec:conclusion}
We have introduced PRBO, a new adaptive trading algorithm for solving the Nonstationary Continuum-Armed Bandit (NCAB) problem by Bayesian optimization and a ``bandit-over-bandit" framework. In a series of empirical simulations, PRBO was compared against PRSH, a reference trading algorithm from the literature. Across a variety of market conditions, PRBO was shown to generate significantly more profit than PRSH, despite having fewer tunable parameters. 
We present this as strong evidence that PRBO is a novel contribution to the field of agent-based computational economics and financial markets simulation. In the wider context, we also present this work as evidence of the potential value of framing problems in finance as NCAB problems, and proposing solutions inspired by the NCAB literature. 

However, the work has some limitations. In particular, we assume that similar parameters produce similar payoffs and the change in the payoff distribution is smooth in time. In future, we will perform variational analysis to better understand the rate of change of payoff distributions. We will also attempt to improve the model by using deep learning approaches to estimate the kernel function in Gaussian processes. Finally, we will evaluate PRBO in more complex markets with time-varying supply and demand, and explore the coevolutionary dynamics of markets containing populations of co-adaptive agents.

%Also, mentioning the managerial implications of the model would make the work more practically relevant.
In a practical application scenario, the model could be used to adapt the parameters of an automated trading system in real time. While the current best-performing parameter set are used for live trading, in parallel an offline simulation environment using live market data feeds is used to continuously update payoff distributions. When new best parameters are identified, the live system is immediately updated with the new best parameter set.

% \section{Funding}
% All sources of funding for the research reported should be declared. The role of the funding body in the design of the study and collection, analysis, and interpretation of data and in writing the manuscript should be declared. Please use \href{http://www.crossref.org/fundingdata/}{FundRef} to report funding sources and include the award/grant number, and the name of the Principal Investigator of the grant. 

% \section{Acknowledgements}
% Please acknowledge anyone who contributed towards the article who does not meet the criteria for authorship including anyone who provided professional writing services or materials. Authors should obtain permission to acknowledge from all those mentioned in the Acknowledgements section. If you do not have anyone to acknowledge, please remove this section.

%%%%%%%%%%%%%%
%% References
% \bibliographystyle{apa}

%% Specify your .bib file name here, without the extension
\bibliography{paper-refs}

\section*{Appendix: Statistical Tests for Normality}

For completeness, here we present evidence that profit distributions are approximately normally distributed. This enables us to safely use the Z-test for statistical significance testing.

\begin{table*}[tb]
    \caption{K-S test result (p-value) of the profits made by PRSH traders under trending market. All profits are approximately normally distributed.}\label{tab:ks-trending}
    \centering
    \begin{tabular}{c|cccc|cccc|cccc}
    \toprule[1.5pt]%第一条粗线
        M & \multicolumn{4}{c|}{m1}  & \multicolumn{4}{c|}{m2} & \multicolumn{4}{c}{m3} \\ \Xhline{1.2pt}%第一条粗线
        \diagbox{K}{V} & 32 & 64 & 128 & 256 & 32 & 64 & 128 & 256 & 32 & 64 & 128 & 256 \\ \Xhline{1.2pt}
        2 & 0.61 & 0.38 & 0.98 & 0.84 & 0.85 & 0.54 & 0.42 & 0.73 & 0.7 & 0.96 & 0.25 & 0.97 \\ \hline 4 & 0.96 & 0.61 & 0.9 & 0.98 & 0.92 & 0.76 & 0.66 & 0.99 & 0.81 & 0.95 & 0.97 & 0.99 \\ \hline 6 & 0.85 & 0.54 & 0.85 & 0.84 & 0.93 & 0.2 & 1.0 & 0.76 & 0.99 & 0.95 & 0.72 & 0.94 \\ \hline 8 & 0.73 & 0.94 & 0.98 & 1.0 & 0.9 & 0.95 & 0.73 & 0.49 & 0.51 & 1.0 & 0.7 & 0.58 \\ \hline 10 & 0.55 & 0.8 & 0.55 & 0.4 & 0.84 & 0.86 & 1.0 & 0.42 & 0.98 & 0.24 & 0.99 & 0.5 \\ \hline 12 & 0.99 & 0.86 & 0.31 & 0.62 & 0.9 & 0.34 & 0.58 & 0.88 & 0.92 & 0.83 & 0.84 & 0.64 \\ \hline 14 & 0.78 & 0.98 & 0.63 & 1.0 & 0.68 & 0.98 & 0.41 & 0.61 & 0.87 & 0.97 & 0.72 & 0.65 \\ \hline 16 & 0.95 & 0.85 & 0.97 & 0.83 & 0.61 & 0.97 & 0.94 & 0.97 & 0.88 & 0.99 & 0.39 & 0.97 \\ \bottomrule[1.5pt]
    \end{tabular}
\end{table*}

\begin{table*}[tb]
    \caption{K-S test result (p-value) of the profits made by PRSH traders under flat market. All profits are approximately normally distributed.}\label{tab:ks-trendless}
    \centering
    \begin{tabular}{c|cccc|cccc|cccc}
    \toprule[1.5pt]%第一条粗线
        M & \multicolumn{4}{c|}{m1}  & \multicolumn{4}{c|}{m2} & \multicolumn{4}{c}{m3} \\ \Xhline{1.2pt}%第一条粗线
        \diagbox{K}{V} & 32 & 64 & 128 & 256 & 32 & 64 & 128 & 256 & 32 & 64 & 128 & 256 \\ \Xhline{1.2pt}
        2 & 0.97 & 0.82 & 0.24 & 0.86 & 0.38 & 0.32 & 0.88 & 0.97 & 0.98 & 0.78 & 0.8 & 0.74 \\ \hline 4 & 0.37 & 0.89 & 0.76 & 0.91 & 0.99 & 0.99 & 0.91 & 0.97 & 0.62 & 0.85 & 0.98 & 0.87 \\ \hline 6 & 0.83 & 0.94 & 1.0 & 0.53 & 0.88 & 0.92 & 0.58 & 0.98 & 0.82 & 0.78 & 0.63 & 0.99 \\ \hline 8 & 0.61 & 0.27 & 0.51 & 0.97 & 0.32 & 0.7 & 0.96 & 0.7 & 0.94 & 0.82 & 0.67 & 0.24 \\ \hline 10 & 0.97 & 0.99 & 0.94 & 0.99 & 0.54 & 0.94 & 0.74 & 0.72 & 0.14 & 0.51 & 0.95 & 0.99 \\ \hline 12 & 0.99 & 0.74 & 0.17 & 0.63 & 0.45 & 0.95 & 0.97 & 0.51 & 0.99 & 0.99 & 0.99 & 0.73 \\ \hline 14 & 0.52 & 0.88 & 0.52 & 0.67 & 0.89 & 0.99 & 0.24 & 0.99 & 0.5 & 0.71 & 0.91 & 0.99 \\ \hline 16 & 1.0 & 0.83 & 0.51 & 0.75 & 0.96 & 0.68 & 0.93 & 0.95 & 0.99 & 0.77 & 0.32 & 0.98 \\ \bottomrule[1.5pt]
    \end{tabular}
\end{table*}

\begin{table*}[t!]
    \caption{K-S test result (p-value) of PRBO profits in trending market and flat market. All profits are approximately normally distributed.}\label{tab:p-values}
    \centering
    \begin{tabular}{c|cccc|cccc}
    \toprule[1.5pt]%第一条粗线
        Market & \multicolumn{4}{c|}{Trending} & \multicolumn{4}{c}{Flat} \\ \cline{1-9}
        \diagbox{K}{V} & 32 & 64 & 128 & 256 & 32 & 64 & 128 & 256 \\ \Xhline{1.2pt}
        2 & 0.58 & 0.33 & 0.47 & 0.82 & 0.96 & 0.95 & 0.92 & 0.89 \\ \hline 
        3 & 0.69 & 1.0 & 0.96 & 0.54 & 0.41 & 0.47 & 0.54 & 0.89 \\ \hline 
        4 & 0.45 & 0.91 & 0.97 & 0.99 & 0.61 & 0.88 & 0.85 & 0.26 \\ \bottomrule[1.5pt]
    \end{tabular}
\end{table*}

\subsection*{PRSH Hyperparameter Exploration}
\label{app:PRSH}
We first test the normality of $X(e,k,v,m)$ using the Kolmogorov–Smirnov (K-S) test with hypotheses: 
\begin{itemize}
\item $\mathbb{H}_0: X(e,k,v,m)$ conforms to the normal distribution. 
\item $\mathbb{H}_1: X(e,k,v,m)$ is not normally distributed. 
\end{itemize} 
\noindent If we cannot reject the null hypothesis $\mathbb{H}_0$, then $X(e,k,v,m)$ conforms to the normal distribution and we perform the Z-test to ascertain whether $\mathbb{E}[X(e,k^*,v^*,m^*)]>\mathbb{E}[X(e,k,v,m)],\forall k,v,m$. We will perform Z-test on each $\bm{x_{e,k,v,m}}$ that we have obtained one-by-one with $\bm{x_{e,k^*,v^*,m^*}}$. The hypothesis of the Z-test is: \begin{itemize}
\item $\mathbb{H}_0: \mathbb{E}[X(e,k^*,v^*,m^*)]\leq\mathbb{E}[X(e,k,v,m)]$
\item $\mathbb{H}_1: \mathbb{E}[X(e,k^*,v^*,m^*)]>\mathbb{E}[X(e,k,v,m)]$
\end{itemize} If at a specific significance level, for some combinations of $\{k,v,m\}$, we cannot reject $\mathbb{H}_0$, then record those $\{k,v,m\}$. Eventually all recorded $\{k,v,m\}$ together with $\{k^*,v^*,m^*\}$ will form $\{K_X^*(e),V_X^*(e),M_X^*(e)\}$. 

Table~\ref{tab:ks-trending} and Table~\ref{tab:ks-trendless} show the K-S test p-values of the profit made by PRSH traders under trending market and flat market, respectively. All p-values exceed 0.05, so we accept that all profits are approximately normally distributed. Therefore, we are able to use Z-test for statistical significance testing.

\subsection*{PRBO Hyperparameter Exploration}
\label{app:PRBO}

We test the normality of $Y(e,k,v)$ using the K-S test, with:
\begin{itemize}
\item $\mathbb{H}_0: Y(e,k,v)$ conforms to the normal distribution. 
\item $\mathbb{H}_1: Y(e,k,v)$ is not normally distributed. 
\end{itemize}
\noindent If $Y(e,k,v)$ conforms to the normal distribution we can then perform the Z-test to ascertain whether $\mathbb{E}[Y(e,k^*,v^*)]>\mathbb{E}[Y(e,k,v)],\forall k,v$. We will perform Z-test on each $\bm{y_{e,k,v}}$ that we have examined one-by-one with $\bm{y_{e,k^*,v^*}}$. The hypothesis of the Z-test is: \begin{itemize}
\item $\mathbb{H}_0: \mathbb{E}[Y(e,k^*,v^*)]\leq\mathbb{E}[Y(e,k,v)]$
\item $\mathbb{H}_1: \mathbb{E}[Y(e,k^*,v^*)]>\mathbb{E}[Y(e,k,v)]$
\end{itemize}If at a specific significance level, for some combinations of $\{k,v\}$, we cannot reject $\mathbb{H}_0$, then record those $\{k,v\}$. Eventually all recorded $\{k,v\}$ together with $\{k^*,v^*\}$ will form $\{K_Y^*(e),V_Y^*(e)\}$. 

\balance
Table~\ref{tab:p-values} shows the K-S test p-values of the profit made by PRBO traders under trending market and flat market, respectively. All p-values exceed 0.05, so we accept that all profits are approximately normally distributed. Therefore, we are able to use Z-test for statistical significance testing.

\subsection*{PRBO vs PRSH: Normality Testing}
\label{app:PRBOvsPRSH}
We test the normality of $D(e)$ using the K-S test: 
\begin{itemize}
\item $\mathbb{H}_0: D(e)$ conforms to the normal distribution. 
\item $\mathbb{H}_1: D(e)$ is not normally distributed. 
\end{itemize}
\noindent If $D(e)$ conforms to the normal distribution we can then perform Z-test to test whether $\mathbb{E}[D(e)]>0$. The hypothesis of the Z-test is: \begin{itemize}
\item $\mathbb{H}_0: \mathbb{E}[D(e)]\leq 0$
\item $\mathbb{H}_1: \mathbb{E}[D(e)]> 0$
\end{itemize}If we can reject the null hypothesis $\mathbb{H}_0$ at a specific significance level, we can accept that PRBO statistically outperforms PRSH. 

Table~\ref{statistics} shows the K-S test p-values of $D$ under trending market and flat market. Both p-values exceed 0.05, so we accept that $D$ is approximately normally distributed. Therefore, we are able to use Z-test for statistical significance testing.

%% End of the document
\end{document}